\documentclass[12pt]{iopart}

\usepackage{iopams}
\usepackage{bm}
\usepackage{graphicx}
\usepackage{color}

\begin{document}

\newcommand{\beq}{\begin{equation}}
\newcommand{\eeq}{\end{equation}}

\newcommand{\REV}[1]{\textbf{\color{red}#1}}
\newcommand{\BLUE}[1]{\textbf{\color{blue}#1}}
\newcommand{\GREEN}[1]{\textbf{\color{green}#1}}

\title[Entanglement in QPTs]{Multipartite entanglement characterization of a quantum phase transition}

\author{G. Costantini$^{1,2}$, P. Facchi $^{3,2}$, G. Florio$^{1,2}$, S.
Pascazio$^{1,2}$}

\address{$^1$Dipartimento di Fisica, Universit\`a di Bari, I-70126 Bari, Italy}
\address{$^2$Istituto Nazionale di Fisica Nucleare, Sezione di Bari, I-70126 Bari, Italy}
\address{$^3$Dipartimento di Matematica, Universit\`a di Bari, I-70125 Bari, Italy}

\begin{abstract}
A probability density characterization of multipartite entanglement
is tested on the one-dimensional quantum Ising model in a transverse
field. The average and second moment of the probability distribution
are numerically shown to be good indicators of the quantum phase
transition. We comment on multipartite entanglement generation at a
quantum phase transition.
\end{abstract}


\section{Introduction}

Quantum phase transitions are characterized by nonanalyticity in
the properties of the states of a physical system \cite{sachdev}.
They differ from classical phase transitions in that they occur at
zero temperature and are therefore driven by quantum (rather than
thermal) fluctuations.

The research of the last few years has unearthed remarkable links
between quantum phase transitions (QPTs) and entanglement
\cite{QPT1,QPT2,QPT3,QPT4}. The study of these inherently quantum
phenomena has mainly focused on bipartite entanglement, by using the
entropy of entanglement \cite{woot}, i.e.\ the von Neumann entropy
of one part of the total system in the ground state. Notwithstanding
the large amount of knowledge accumulated, the properties of the
\emph{multipartite} entanglement of the ground state at the critical
points of a QPT are not clear yet. This is also due to the lack of a
unique definition of multipartite entanglement \cite{druss}.
Different definitions tend indeed to focus on different aspects of
the problem, capturing different features of the phenomenon
\cite{multipart}, that do not necessarily agree with each other.
This is basically due to the fact that, as the size of the system
increases, the number of measures (i.e.\ real numbers) needed to
quantify multipartite entanglement grows exponentially. For all
these reasons, the quantification of multipartite entanglement is an
open and very challenging problem.

In the study of a QPT the above-mentioned problems are of great
importance. The evaluation of entanglement measures bears serious
computational difficulties, because the ground states involve
exponentially many coefficients. The issue is therefore to
understand how to characterize entanglement, e.g.\ by identifying
one key property that can summarize its multipartite features. Our
strategy will be to look at the probability density function of the
purity of a subsystem over all bipartitions of the total system. The
average of this function will determine the amount of global
entanglement in the system, while the variance will measure how well
such entanglement is distributed: a smaller variance will correspond
to a larger insensitivity to the choice of the bipartition and,
therefore, will witness if entanglement is really multipartite.

This approach, introduced in \cite{FFP}, makes use of statistical
information on the state and extends in a natural way the techniques
used for the bipartite entanglement. It is interesting to notice
that the idea that complicated phenomena cannot be ``summarized" by
a single (or a few) number(s) was already proposed in the context of
complex systems \cite{parisi} and has been also considered in
relation to quantum entanglement \cite{MMSZ}. We applied our
characterization of multipartite entanglement to a large class of
random states \cite{aaa,SC}, obtaining sensible results
\cite{FFP,FFP2}.

In this article we will characterize in a similar way the
multipartite entanglement of the (finite) Ising model in a
transverse field. Our numerical results will corroborate previous
findings and yield new details about the structure of quantum
correlations near the quantum critical point.

\section{Probability density function characterization
of multipartite entanglement}

We shall focus on a collection of $n$ qubits and consider a
partition in two subsystems $A$ and $B$, made up of $n_A$ and
$n_B$ qubits ($n_A+n_B=n$), respectively. For definiteness we
assume $n_A \le n_B$. The total Hilbert space is the tensor
product $\mathcal{H}=\mathcal{H}_A\otimes\mathcal{H}_B$ with
dimensions $\dim \mathcal{H}_A=N_A=2^{n_A}$, $\dim
\mathcal{H}_B=N_B=2^{n_B}$ and  $\dim \mathcal{H}=N=N_AN_B=2^n$.

We shall consider pure states
\begin{equation}
|\psi\rangle = \sum_{k=0}^{N-1} z_k |k\rangle =
\sum_{j_A=0}^{N_A-1} \sum_{l_B=0}^{N_B-1} z_{j_A l_B}
|j_A\rangle\otimes|l_B\rangle , \label{eq:genrandomx}
\end{equation}
where the last expression is adapted to the bipartition:
$|k\rangle=|j_A\rangle\otimes|l_B\rangle$, with a bijection between
$k$ and $(j_A,l_B)$. Think for example of the binary expression of
an integer $k$ in terms of the binary expression of $(j_A, l_B)$. We
define the purity (linear entropy) of the subsystem
\begin{equation}\label{eq:NABdefs}
\pi_{AB}(|\psi\rangle)=\Tr_{A}
\rho_A^2,   \qquad \rho_A=\Tr_B \rho, \qquad
\rho=|\psi\rangle\langle\psi|,
\end{equation}
$\Tr_A$ ($\Tr_B$) being the partial trace over subsystem $A$ ($B$),
and take as a measure of the bipartite entanglement between $A$ and
$B$ the participation number
\begin{equation}\label{eq:NAB}
N_{AB}=\pi_{AB}^{-1},
\end{equation}
that measures the effective rank of the matrix $\rho_A$, namely
the effective Schmidt number \cite{eberly}. The quantity $n_{AB}=
\log_2 N_{AB}$ represents the effective number of entangled
qubits, given the bipartition (pictorially, the number of bipartite
entanglement ``links" that are ``severed" when the system is
bipartitioned). By plugging (\ref{eq:genrandomx}) into
(\ref{eq:NABdefs}) one gets
\begin{equation}\label{eq:piAB}
N_{AB}(|\psi\rangle) =
\left(\sum_{j,j'=0}^{N_A-1}\,\sum_{l,l'=0}^{N_B-1}z_{j l} \bar
z_{j' l} z_{j' l'} \bar z_{j l'}\right)^{-1}.
\end{equation}
This is the key formula of our numerical investigation.

Clearly, the quantity $N_{AB}$ will depend on the bipartition, as in
general entanglement will be distributed in a different way among
all possible bipartitions. We are pursuing the idea that the density
function $p(N_{AB})$ of $N_{AB}$ yields information about
\emph{multipartite} entanglement \cite{FFP}. We note that
\begin{equation}\label{eq:propNAB}
1\leq N_{AB}=N_{BA} \leq N_A (\leq N_B),
\end{equation}
where the maximum (minimum) value is obtained for a completely
mixed (pure) state $\rho_A$. Therefore, a larger value of $N_{AB}$
corresponds to a more entangled bipartition $(A,B)$. Incidentally,
we notice that the maximum possible bipartite entanglement
$N_{AB}^{\mathrm{max}} = N_A^{\mathrm{max}}= 2^{[n/2]}$ can be
attained only for a balanced bipartition, i.e.\ when $n_A = [n/2]$
(and $n_B=[(n+1)/2]$), where $[x]$ is the integer part of the real
$x$, that is the largest integer not exceeding $x$. We emphasize
that the use of the inverse purity (linear entropy) (\ref{eq:NAB})
is only motivated by simplicity. Any other measure of bipartite
entanglement, such as the entropy (or any Tsallis entropy
\cite{tsallis}) would yield similar results.

\section{Entanglement distribution: critical Ising chain in a transverse field.}

We now apply the characterization of multipartite entanglement to
the quantum Ising chain in a transverse field, described by the
Hamiltonian
\begin{equation}\label{Isingwithtransverse}
H=-g\sum_{i=1}^{n-1}\sigma_i^z \sigma_{i+1}^z
-(1-g)\sum_{i=1}^{n}\sigma_i^x + \epsilon\sum_{i=1}^{n}\sigma_i^z
\end{equation}
(with open boundary conditions, $\sigma$ being the Pauli matrices).
Notice that we added a (small, site independent) longitudinal field
$\epsilon$. If $\epsilon=0$, it is known from conformal field theory
\cite{wilczek} and numerical simulations based on accurate
analytical expressions
\cite{QPT2} that at the critical point $g=g_{\rm c}=1/2$ the
entanglement entropy
\begin{equation}
S_{AB}=-\Tr_{A}(\rho_A \log_2 \rho_A)
\end{equation}
diverges with a logarithmic law
\begin{equation}
\label{eq:divergenceentropy} S_{AB} \sim \frac{1}{6}\log_2 \ell.
\end{equation}
Here entanglement is evaluated by considering a block $A$ of
\emph{contiguous} spins whose length $\ell$ is less than one half the
total length $n$ of the chain. Due to (approximate) translation
invariance, in our approach this is equivalent to considering the
average entanglement over a subset of the bipartitions of the system
(that tend to be balanced when $\ell$ tends to $n/2$).

\subsection{A typical distribution}

We intend to evaluate the distribution of bipartite entanglement
over all balanced bipartitions and, therefore, the multipartite
entanglement. Here and in the whole article, the Hamiltonian will be
exactly diagonalized in order to obtain the ground state, then
$N_{AB}$ will be explicitly evaluated as a function of $g$ and its
distribution plotted. The results are exact, but the quantum
simulation time consuming and for this reason $n$ cannot be too
large.

\begin{figure}
\begin{center}
\includegraphics[width=0.95 \textwidth]{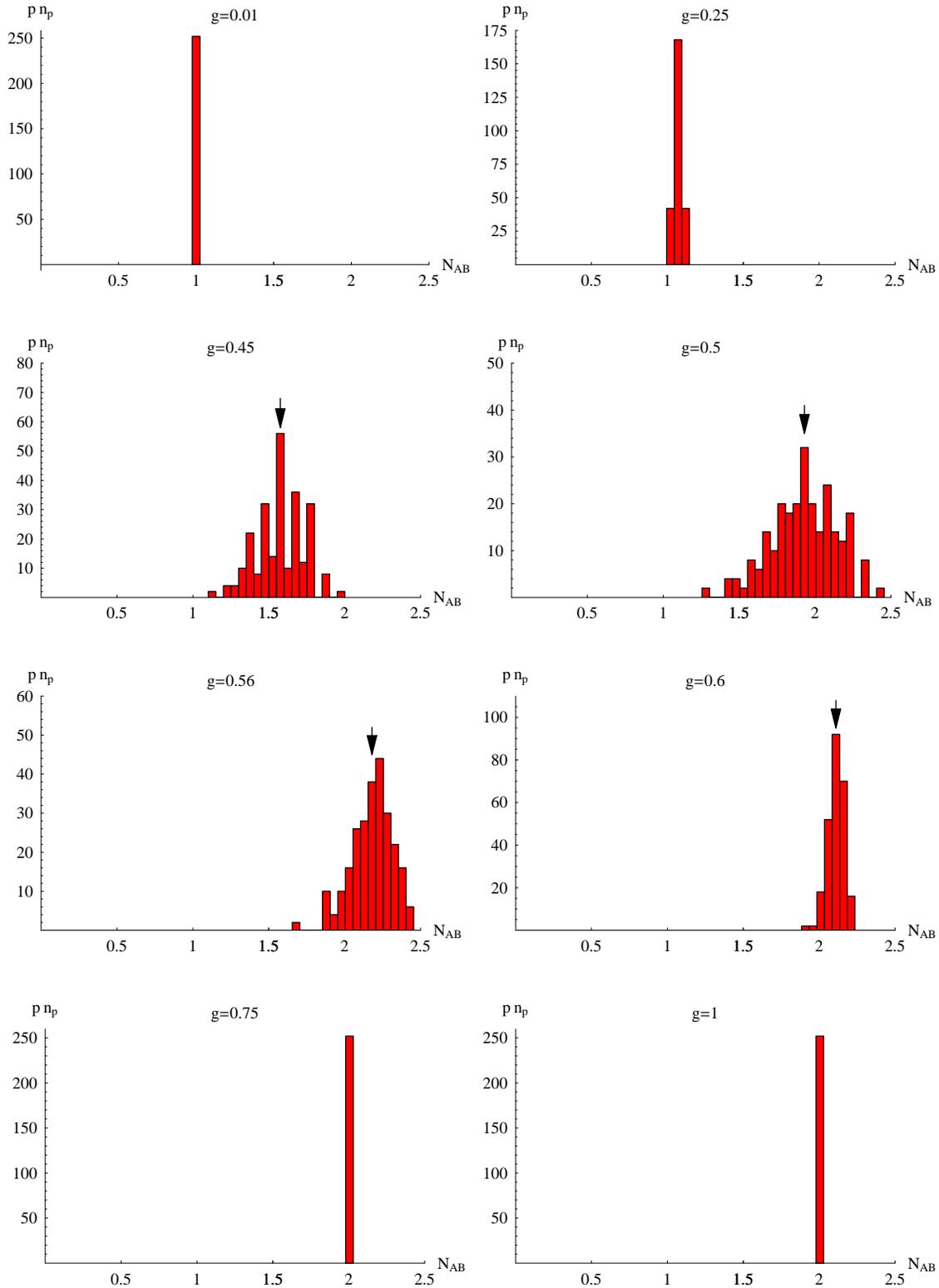}
\caption{Distribution function of the participation number
$N_{AB}$ over all balanced bipartitions for the Hamiltonian
(\ref{Isingwithtransverse}) when $\epsilon =0$ and $n=10$. The
distribution is always bell-shaped. Its width is maximum at $g=0.5$,
while its average entanglement (indicated by a black arrow) is
maximum at $g=0.56$. Notice the different scales on the ordinates.
The number of balanced bipartitions is $n_p=\left(\!\!\!
                                               \begin{array}{c}
                                                 n \cr
                                                 [n/2]
                                               \end{array}
                                             \!\!\!\right)=\left(\!\!\!
                                               \begin{array}{c}
                                                 10 \cr
                                                 5
                                               \end{array}
                                             \!\!\!\right)
=252$. } \label{fig:probframes}
\end{center}
\end{figure}

The distribution of the participation number $N_{AB}$ as $g$ varies,
for $n=10$ qubits and $\epsilon=0$, is shown in Fig.\
\ref{fig:probframes}.
We notice that the distribution is always well-behaved and
bell-shaped, being practically a $\delta$ function for $g\leq 0.1$
and $g\geq 0.75$. For this reason, one can get a satisfactory
characterization of multipartite entanglement by looking at its mean
value and width
\begin{equation}
\mu=\langle
N_{AB}\rangle,\quad \sigma^2=\langle{(N_{AB}-\mu)}^2\rangle,
\end{equation}
where the average $\langle \cdots \rangle$ is evaluated over all
balanced bipartitions. We recall that $\mu$ defines the amount of
entanglement while the inverse width $\sigma^{-1}$ describes how
fairly such entanglement is distributed. We notice that the width
$\sigma$ is maximum at $g=0.5$, while the average entanglement $\mu$
is maximum at $g=0.56$. Observe that no singularities can be
expected for a number of spins as small as $n=10$, yet the behavior
of both quantities clearly foreruns the quantum phase transition at
$g=g_c=1/2$. In this sense, both $\sigma$ and $\mu$ appear to be
good indicators of the QPT.
\begin{figure}
\begin{center}
\includegraphics[width=1.0 \textwidth]{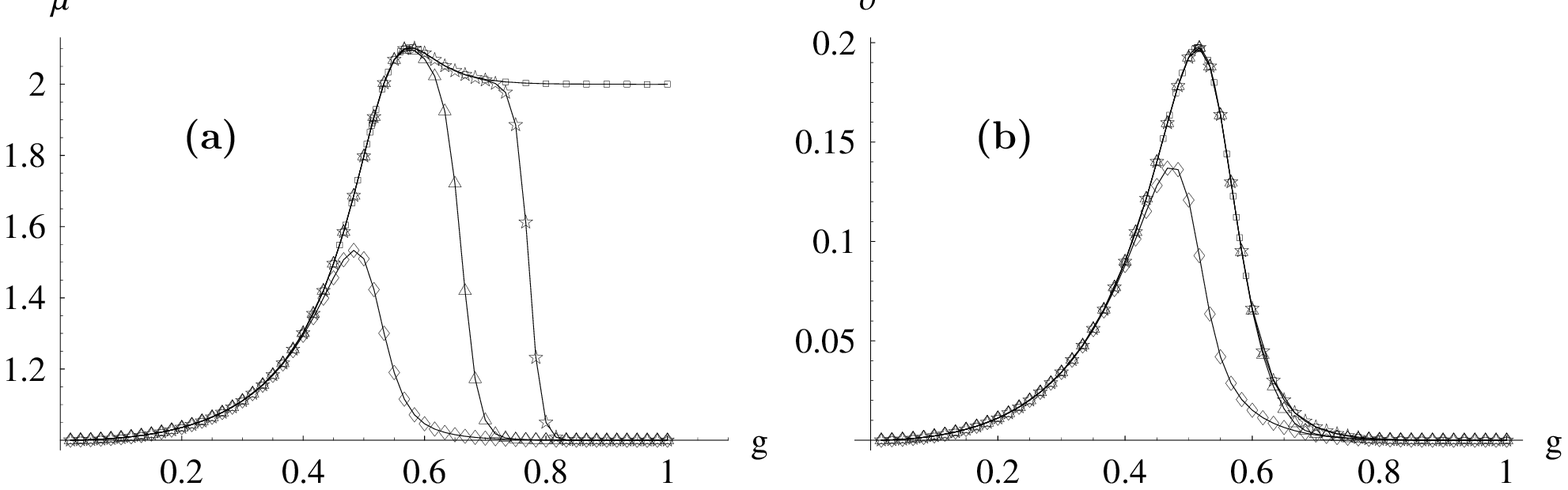}
\caption{(a) Average $\mu$ and (b) standard deviation $\sigma$ of
$N_{AB}$ over all balanced bipartitions for $n=9$ sites in the 1-D
quantum Ising model in a transverse field of strength $1-g$ and a
small longitudinal field of strength $\epsilon$. Squares:
$\epsilon=0$; stars: $\epsilon=10^{-6}$; triangles:
$\epsilon=10^{-4}$; diamonds: $\epsilon=10^{-2}$.
}\label{fig:musigmanoise}
\end{center}
\end{figure}

\subsection{Average and width}

Let us consider the full Hamiltonian (\ref{Isingwithtransverse})
when the longitudinal perturbing field is small. In Fig.\
\ref{fig:musigmanoise} we plot $\mu$ and $\sigma$, respectively,
vs $g$ for the ground state of the Hamiltonian
(\ref{Isingwithtransverse}), when $n=9$, for different values of
$\epsilon$ (ranging from $0$ to $10^{-2}$). We notice a very
different behavior of the two quantities. The average $\mu$ is
very sensitive to the longitudinal perturbation. In the region
$g\simeq 1$, where the ground state is approximately a GHZ state
when $\epsilon=0$, the average entanglement is strongly reduced
even for a very small value of $\epsilon$ ($\simeq 10^{-6}$). This
is basically due to the fact that the superposition $|{\rm all \;
spins \; up}\rangle + |{\rm all \; spins \; down}\rangle$
(yielding $\mu=2$) is very fragile and the ground state collapses
in one of the two (degenerate) classical ground states (yielding
$\mu=1$), the $Z_2$ symmetry being broken. On the other hand, near
the maximum, $\mu$ is more robust and a larger perturbation
($\epsilon=10^{-2}$) is required to counter larger values of
$(1-g)$ and modify the behavior of $\mu$.

The behavior of $\sigma$ is different. When $\epsilon \lesssim
10^{-2}$ the curves are not modified by the presence of the
longitudinal field. This is due to the fact that in the region where
$\mu$ is reduced by the presence of $\epsilon$, $\sigma$ is already
near to 0 (a GHZ state has $\sigma=0$ because it is invariant for
permutation of the qubits, see \cite{FFP}). Of course, a
sufficiently large value of $\epsilon$ affects also $\sigma$,
reducing it (but not modifying the shape of $\sigma(g)$).

We shall now focus on the critical region. It would be tempting to
take a small value of $\epsilon$ (say, $\epsilon =10^{-6}$) in
order to get rid of the spurious residual entanglement at $g\simeq
1$ (and obtain a bell-shaped function for $\mu$---as well as for
$\sigma$). However, since we aim at a precise determination of the
coordinates of the maximum, which is unaffected by small values of
$\epsilon$, we decided to work with $\epsilon=0$.

\subsection{Purely transverse Ising chain}

\begin{figure}
\begin{center}
\includegraphics[width=1.0 \textwidth]{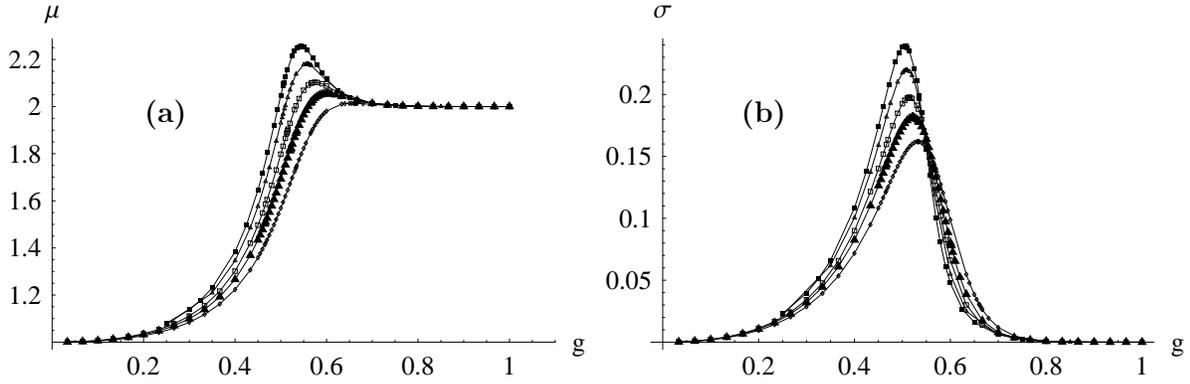}
\caption{(a) Average $\mu$ and (b) standard deviation $\sigma$ of
$N_{AB}$ over all balanced bipartitions (from $n=7$ to $11$ sites)
for the purely transverse 1-D ising chain. Full squares: 11 sites;
open triangles: 10 sites; open squares: 9 sites; full triangles: 8
sites; open diamonds: 7 sites. $\mu$ can be viewed as a measure of
the average multipartite entanglement, while $\sigma^{-1}$ can be
viewed as a measure of how fairly this entanglement is shared. Both
$\mu$ and $\sigma$ are good indicators of the QPT that takes place
at $g=0.5$. Interestingly, $\sigma_{\mathrm{max}}$ precedes
$\mu_{\mathrm{max}}$.
\label{fig:musigma}}
\end{center}
\end{figure}

In Fig.\ \ref{fig:musigma} we evaluate the average and standard
deviation for $\epsilon=0$ (purely transverse) Ising chains of
increasing size (from 7 to 11 sites). In Fig.\
\ref{fig:musigma}(a) we distinguish different zones. For
$g=0$ the ground state (gs) is factorized and $\mu=1$. If $g\simeq
1$ the gs is approximately a GHZ state (a combination of the gs's of
the classical Hamiltonian). The most interesting region is around
the value $g=0.5$, where for an increasing number of qubits there is
a more and more pronounced peak of $\mu$. This is in qualitative
agreement with other results obtained using the entropy of
entanglement.

The width of the distribution of $N_{AB}$ versus $g$ is shown in
Fig.\ \ref{fig:musigma}(b). We will comment later on the behavior of
this quantity, that yields useful additional information about the
structure and generation of multipartite entanglement (information
that would not be available for an entanglement measure constituted
by a single number). Also in this case we can distinguish several
regions in the plot. Moreover, the coupling $g$ corresponding to the
peak of $\sigma$ (that we denote $\sigma_{\rm max}$), does not
coincide with that corresponding to the peak of $\mu$ (that we
denote $\mu_{\rm max}$):
\begin{equation}
g\left(\sigma_{\rm max}\right)<g\left(\mu_{\rm max}\right).
\end{equation}
In other words, for a finite spin chain, the width of the
distribution is not maximum when the amount of entanglement is
maximum.

\begin{figure}
\begin{center}
\includegraphics[width=1.0 \textwidth]{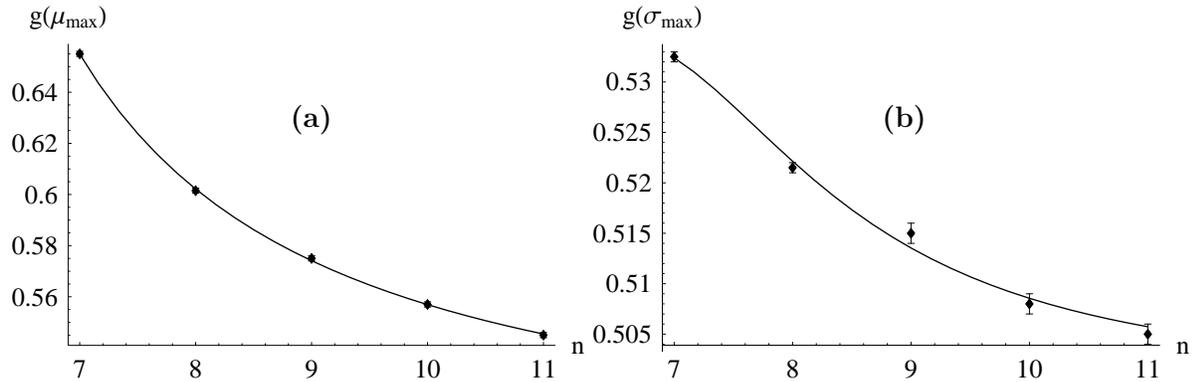}
\caption{Coupling constant $g$ corresponding to (a) $\mu_{\rm
max}$ and (b) $\sigma_{\rm max}$ versus $n$. Notice that
$g(\sigma_{\rm max})<g(\mu_{\rm max})$ (at fixed $n$) and that
already for small $n(=7)$, $g(\sigma_{\rm max})$ differs from
$g_c=1/2$ only by a few percent. The error bars (one standard
deviation) are explicitly indicated. \label{fig:gmugsigma}}
\end{center}
\end{figure}

We notice that, by increasing $n$, both maxima are shifted towards
the center of the plot $g \to g_{\rm c}=0.5$. In Fig.\
\ref{fig:gmugsigma}(a) we plot the values of the coupling constant
$g$ at $\mu_{\rm max}$ versus the number of sites $n$. The
numerical result can be fitted with the (arbitrary) function
\begin{equation}
\label{eq:fitgn} g(\mu_{\rm max})=0.5 + \frac{5.43}{n^2
+3.09\,n -35.59} \stackrel{n \to \infty}{\longrightarrow} 0.5=g_{c}
.
\end{equation}
The plot of $g(\sigma_{\rm max})$ versus $n$ is shown in Fig.\
\ref{fig:gmugsigma}(b), the fit being
\begin{equation}
\label{eq:gsigmamaxn} g\left(\sigma_{\rm max}\right)=0.5 +
\frac{0.14}{n^2 - 13.01\,n +46.39}  \stackrel{n \to \infty}{\longrightarrow}
0.5=g_{c}.
\end{equation}
Notice that the fit (\ref{eq:fitgn}) is very accurate, while
(\ref{eq:gsigmamaxn}) is valid within one standard deviation
(namely a few percent), as can be seen in Fig.\
\ref{fig:gmugsigma}.
From Fig.\ \ref{fig:gmugsigma} and  Eqs.\
(\ref{eq:fitgn})-(\ref{eq:gsigmamaxn}) one can argue that the amount
of entanglement (the mean of the distribution) and the maximum width
of the distribution of bipartite entanglement can detect, in the
limit of large $n$, the QPT.


\begin{figure}
\begin{center}
\includegraphics[width=1.0 \textwidth]{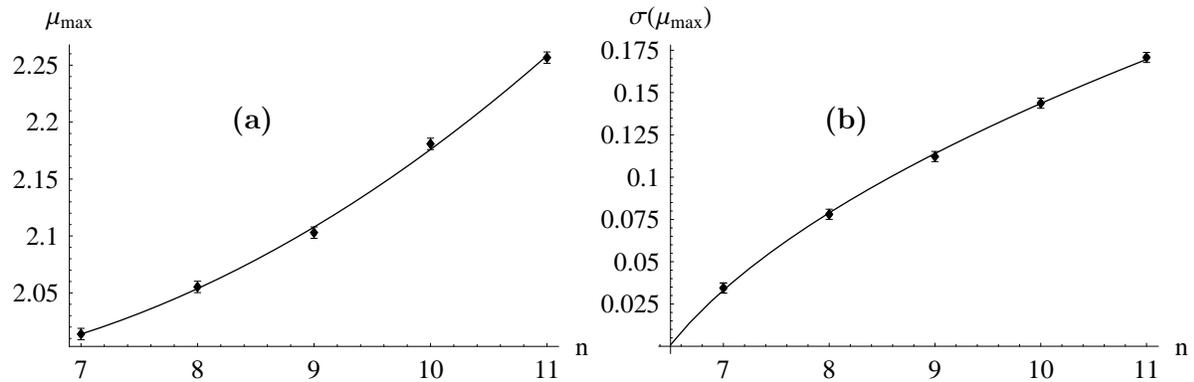}
\caption{(a) Entanglement $\mu_{\rm max}$ and (b) standard
deviation at the maximum entanglement $\sigma(\mu_{\rm max})$ vs
$n$. The error bars (one standard deviation) are explicitly
indicated.} \label{fig:mumaxsigmamumax}
\end{center}
\end{figure}

\begin{figure}
\begin{center}
\includegraphics[width=0.9 \textwidth]{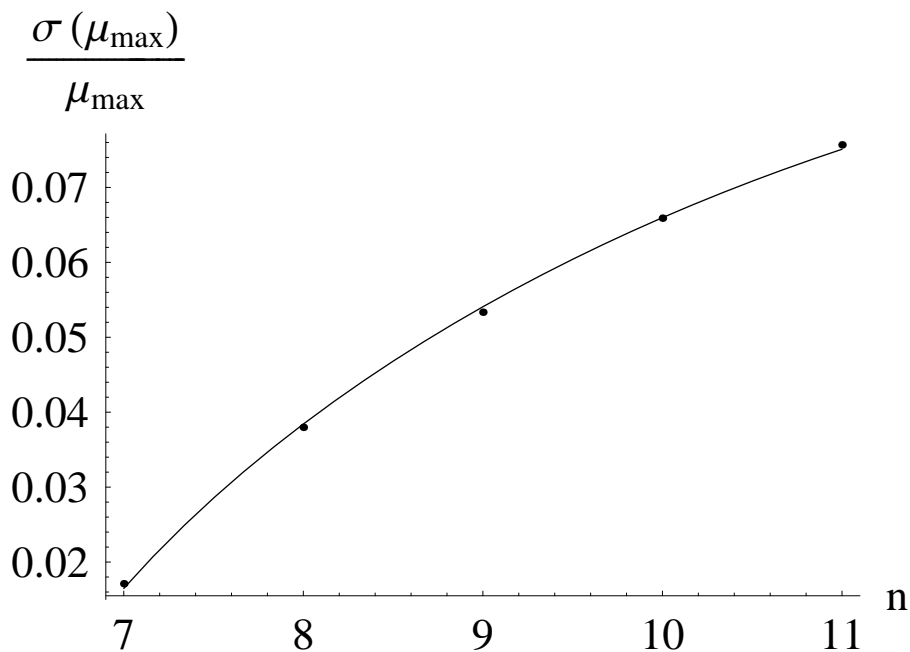}
\caption{Ratio $\sigma(\mu_{\rm max})/\mu_{\rm max}$ vs
$n$.} \label{fig:fracsnabmax}
\end{center}
\end{figure}

We shall henceforth focus on $\mu_{\rm max}$ and $\sigma(\mu_{\rm
max})= \sigma(g(\mu_{\rm max}))$ (the value of $\sigma$ when the
amount of entanglement is maximum), rather than $\sigma_{\rm max}$
(whose behavior is anyway similar). In Fig.\
\ref{fig:mumaxsigmamumax} we plot these quantities vs the number of
spins $n$. They are fitted by (for $n
\ge 6$)
\begin{eqnarray}
\label{eq:fitnabmaxn} \mu_{\rm max}&=&2 +
0.019\,\left(  n -6 \right) +
  0.007\,{\left( n-6 \right) }^2, \\
\sigma(\mu_{\rm max})&=&-0.077 + 0.11\,{\sqrt{ n-6}} .
\label{eq:fitsabmaxn}
\end{eqnarray}
We also evaluate the relative width at maximum entanglement
\beq
\sigma_{\rm rel}=\sigma(\mu_{\rm max})/\mu_{\rm max},
\label{eq:fitsabmaxn1}
\eeq
shown in Fig.\ \ref{fig:fracsnabmax}, that will be useful in the
following discussion. The fitting curve in Fig.\
\ref{fig:fracsnabmax} is not independent, but is rather derived from
Eqs.\ (\ref{eq:fitnabmaxn})-(\ref{eq:fitsabmaxn}):
\beq
\sigma_{\rm rel}= \frac{-0.077 + 0.11\,{\sqrt{ n-6}}}{2 +0.019\,\left(  n -6 \right) +
0.007\,{\left( n-6 \right) }^2}.
\label{eq:fitsabmaxnfit}
\eeq

\section{Discussion}

Both fits (\ref{eq:fitnabmaxn})-(\ref{eq:fitsabmaxn}) imply that the
entanglement indicators $\sigma$ and $\mu$ diverge with $n$ at the
QPT. This conclusion is particularly significant: the amount of
entanglement goes to infinity but so does the width of the
entanglement distribution. In particular, this leads to two possible
scenarios, depending on the behavior of $\sigma_{\rm rel}$ defined
in (\ref{eq:fitsabmaxn1}):
\begin{enumerate}
    \item
    $\sigma_{\rm rel}\stackrel{n\rightarrow\infty}{\longrightarrow}0$.
    In this case the divergence of $\mu_{\rm max}$ is
    stronger than that of $\sigma(\mu_{\rm max})\simeq\sigma_{\mathrm{max}}$.
    This means that at the QPT the entanglement of the ground
    state is macroscopically insensitive to the choice of the
    bipartition. Accordingly, the QPT yields a fair distribution of
    bipartite entanglement and is therefore a good tool for
    generating multipartite entanglement. This conclusion could pave the way towards a
    deeper understanding of the relation among entanglement, QPTs and
    chaotic systems (that are known to generate large amounts of entanglement \cite{SC,entvschaos}).
    \item
    $\sigma_{\rm rel}\stackrel{n\rightarrow\infty}{\longrightarrow}c > 0\,\,(\mbox{eventually}\,\infty)$.
    This situation would have profound consequences
    on our comprehension of the relation between a QPT and the
    generation of multipartite entanglement. In particular, the strong
    divergence of $\sigma(\mu_{\rm max})$ (of order equal to or larger than that of $\mu_{\rm max}$)
    would imply that the distribution of entanglement is \emph{not}
    optimal, inasmuch as it is not fairly shared. This means that the amount of entanglement of non-contiguous spins
    partitions macroscopically differs from that of contiguous
    ones.
\end{enumerate}
Our results, although not conclusive due to the relatively small
value of $n$ reached in our numerical analysis, appear to indicate
that (i) is the most probable scenario: indeed, from Eq.\
(\ref{eq:fitsabmaxnfit}), that in turn is a consequence of Eqs.\
(\ref{eq:fitnabmaxn})-(\ref{eq:fitsabmaxn}), we infer that for
large $n$
\beq
\sigma_{\rm rel} \sim n^{-3/2}.
\label{eq:conclusione}
\eeq
In general, if one assumes that the behavior of $\mu_{\rm max}$
and $\sigma(\mu_{\rm max})$ vs $n$ (and in particular the
convexity of the two curves) does not change for larger $n$, one
can conclude that $\sigma_{\rm rel}$ vanishes for $n\to\infty$.

Another important observation, related to the ``entangling power''
of evolutions \cite{zanardi}, is the following. Although our
numerical results seem to favor the first scenario, namely a well
distributed multipartite entanglement generated by the quantum phase
transition, such entanglement is \emph{not so large}. Indeed, a
\emph{typical} $n$-qubit state is characterized by \cite{FFP}
\begin{equation}\label{typical}
\mu
\propto 2^{n/2}, \qquad \sigma=\mathrm{const},
\end{equation}
namely an exponentially large amount of entanglement, that is also
very well distributed. These typical states are efficiently produced
by a chaotic dynamics \cite{SC,entvschaos}. In general, one observes
a very rapid growth of the (effective) Schmidt number (\ref{eq:NAB})
at the onset of chaos and for all these reasons, quantum chaos is a
much better multipartite entanglement generator than a critical
Ising chain. This conclusion seems to be valid for other spin
Hamiltonians as well. Notice that the entangling power (and/or
entanglement generation) of a QPT is better compared to that of a
chaotic system \cite{entvschaos} (in that they are both obtained by
varying one or more coupling constants), rather than that of a
quantum evolution \cite{zanardi}. On the other hand, unlike in a
chaotic system, in a QPT one focuses on the features of the ground
state.

The entanglement generation at a QPT and the physical features of
this entanglement \cite{grigolin,bolognesi} deserve additional
investigations. The participation number or the entropy of
entanglement (or any other sensible measure) are related to the
global structure of the state. It is therefore reasonable to expect
that many observables might be necessary in order to characterize
multipartite entanglement. The approach we propose \cite{FFP,FFP2},
based on the calculation of the probability density function of
bipartite entanglement, has the advantage of making use of
statistical information on the state of the system and characterizes
multipartite entanglement by extending techniques that are widely
used in the analysis of its bipartite aspects. We have seen that
when the density functions are well behaved and bell-shaped, the
average and second moment of the distribution are good indicators of
the quantum phase transition. These conclusions must be corroborated
by the study of other systems and models displaying quantum phase
transitions, as well as by the analysis of more complex systems
\cite{parisi,MMSZ}. Work is in progress in this direction.

\ack
We thank A. Scardicchio and K. Yuasa for interesting remarks. This
work is partly supported by the European Community through the
Integrated Project EuroSQIP.

\end{document}